\documentstyle[12pt,aps]{revtex}

\widetext
\draft
\tighten
\begin{document}

\preprint{EFUAZ FT-96-29}

\title{Comment on ``Describing Weyl Neutrinos by a Set of Maxwell-like
Equations"  by S. Bruce\thanks{Submitted to ``Nuovo Cimento B".}}

\author{{\bf Valeri V. Dvoeglazov}}

\address{
Escuela de F\'{\i}sica, Universidad Aut\'onoma de Zacatecas \\
Antonio Doval\'{\i} Jaime\, s/n, Zacatecas 98068, ZAC., M\'exico\\
Internet address:  VALERI@CANTERA.REDUAZ.MX}

\date{July 18, 1996}

\maketitle

\bigskip

\begin{abstract}
Results of the work of S. Bruce [{\it Nuovo Cimento}
{\bf 110}B (1995) 115] are compared with those of recent
papers of D. V. Ahluwalia and  myself,
devoted to describing neutral particles of spin $j=1/2$
and $j=1$.
\end{abstract}


\newpage

The main result of ref.~\cite{Bruce} is proving the possibility
of deriving the generalized Maxwell's equations (Eqs. (24)
of the cited paper) from a set of the $j=1/2$ equations for Weyl spinors.
Another important point discussed there is a connection between
parity-even and parity-odd parts of the $CP$ conjugate states. The
consideration is restricted by massless case.

On the other hand in recent papers of D. V. Ahluwalia and
V. V. Dvoeglazov~\cite{NP}
on the basis of ideas of E. Majorana~\cite{Maj} and J. A. McLennan and
K. M. Case~\cite{MLC} the construct for self/anti-self charge conjugate
states has been presented. It permits one to take into account
possible effects of neutrino mass and to explain  origins
of the question of ``missing" right-handed neutrino~\cite{Ziino}.
Type-II self/anti-self charge conjugate spinors
are defined in the momentum representation from the beginning, in the
following way~[2a,Eq.(6)]:
\begin{eqnarray} \lambda(p^\mu)\,\equiv
\pmatrix{ \left ( \zeta_\lambda\,\Theta_{[j]}\right
)\,\phi^\ast_{_L}(p^\mu)\cr \phi_{_L}(p^\mu)} \,\,,\quad
\rho(p^\mu)\,\equiv \pmatrix{ \phi_{_R}(p^\mu)\cr \left (
\zeta_\rho\,\Theta_{[j]}\right )^\ast \,\phi^\ast_{_R}(p^\mu)} \,\,\quad
.\label{sp-dva}
\end{eqnarray}
$\zeta_\lambda$ and $\zeta_\rho$ are the
phase factors that are  fixed by the conditions of  self/anti-self
conjugacy, $\Theta_{[j]}$ is the Wigner time-reversal operator for spin
$j$.  They satisfy the equations\footnote{As we got knowing recently
this set of equations has been proposed in ref.~\cite{Markov} for the
first time.}
\begin{mathletters} \begin{eqnarray} i \gamma^\mu
\partial_\mu \lambda^S (x) - m \rho^A (x) &=& 0 \quad, \label{11}\\ i
\gamma^\mu \partial_\mu \rho^A (x) - m \lambda^S (x) &=& 0 \quad,
\label{12}\\ i \gamma^\mu \partial_\mu \lambda^A (x) + m \rho^S (x) &=&
0\quad, \label{13}\\ i \gamma^\mu \partial_\mu \rho^S (x) + m \lambda^A
(x) &=& 0\quad.  \label{14}
\end{eqnarray}
\end{mathletters}
It was shown
there (cf.~[2b,Eqs. (22)] or~[2c,Eqs.(67-70)])) that type-II spinors
are connected with the Dirac spinors in the Weyl representation $u_\sigma
(p^\mu)$ and $v_\sigma (p^\mu)$ as follows:
\begin{mathletters}
\begin{eqnarray} \lambda^S_{\uparrow} (p^\mu) &=& {1\over 2} \left (
u_{_{+1/2}} (p^\mu) +i u_{_{-1/2}} (p^\mu) - v_{_{+1/2}} (p^\mu) +i
v_{_{-1/2}} (p^\mu)\right )\quad,\\ \lambda^S_{\downarrow} (p^\mu) &=&
{1\over 2} \left ( -i u_{_{+1/2}} (p^\mu) + u_{_{-1/2}} (p^\mu) - i
v_{_{+1/2}} (p^\mu) - v_{_{-1/2}} (p^\mu)\right )\quad,\\
\lambda^A_{\uparrow} (p^\mu) &=& {1\over 2} \left ( u_{_{+1/2}} (p^\mu) -i
u_{_{-1/2}} (p^\mu) - v_{_{+1/2}} (p^\mu) - i v_{_{-1/2}} (p^\mu)\right
)\quad,\\ \lambda^A_{\downarrow} (p^\mu) &=& {1\over 2} \left ( i
u_{_{+1/2}} (p^\mu) + u_{_{-1/2}} (p^\mu) +i v_{_{+1/2}} (p^\mu) -
v_{_{-1/2}} (p^\mu)\right )\quad.
\end{eqnarray} \end{mathletters}
and~[2a,Eqs.(48)]
\begin{mathletters}
\begin{eqnarray}\label{i1}
\rho^{S}_\uparrow (p^\mu) = -i \lambda^A_\downarrow (p^\mu)\quad,\quad
\rho^{A}_\uparrow (p^\mu) = +i \lambda^S_\downarrow (p^\mu)\quad,\\
\label{i2}
\rho^{S}_\downarrow (p^\mu) = + i \lambda^A_\uparrow (p^\mu)\quad,\quad
\rho^{A}_\downarrow (p^\mu) = - i \lambda^S_\uparrow (p^\mu)\quad.
\end{eqnarray}
\end{mathletters}
We assumed that $\phi_{_{L,R}}^{^{+1/2}}
(\overcirc{p}^\mu) = column (1\quad 0)$,\,\, $\phi_{_{L,R}}^{^{-1/2}}
(\overcirc{p}^\mu) = column (0\quad 1)$; in the opposite case we have to
include additional phase factors in the mass terms of Eqs.
(\ref{11}-\ref{14}). They can be fixed if  the theory is implied invariant
with respect to intrinsic parity.

Using identities (\ref{i1},\ref{i2}) and rewriting the equations
(\ref{11}-\ref{14}) in the momentum representation with taking into
account the chiral helicity quantum number~\cite{NP} we are able
to obtain the following equations in the two-component form (phase
factors are restored):
\begin{mathletters}
\begin{eqnarray}\label{111}
\left [p^0 + \bbox{\sigma}\cdot {\bf p} \right ] \phi_{_L}^\uparrow (p^\mu)
-m e^{+i\chi} \Theta \phi_{_L}^{\downarrow\,\,\ast} (p^\mu)
&=&0\quad,\\
\label{112} \left [p^0 - \bbox{\sigma}\cdot {\bf p} \right ]
\Theta \phi_{_L}^{\uparrow \,\,\ast} (p^\mu) + m e^{-i\chi}
\phi_{_L}^{\downarrow} (p^\mu) &=&0\quad,\\
\label{113} \left [p^0 +
\bbox{\sigma}\cdot {\bf p} \right ] \phi_{_L}^{\downarrow} (p^\mu) + m
e^{+i\chi} \Theta \phi_{_L}^{\uparrow\,\,\ast} (p^\mu) &=&0\quad,\\
\label{114} \left [p^0 -
\bbox{\sigma}\cdot {\bf p} \right ] \Theta \phi_{_L}^{\downarrow\,\,\ast}
(p^\mu) - m e^{-i\chi} \phi_{_L}^{\uparrow} (p^\mu) &=&0\quad,
\end{eqnarray}
\end{mathletters}
which answer for the McLenann-Case-Ahluwalia construct. A remarkable
feature of  this set is that it is valid both for positive- and
negative-energy solutions of the equations (\ref{11}-\ref{14}).
The corresponding equations for $\phi_{_R} (p^\mu)$ and $\Theta
\phi_{_R}^\ast (p^\mu)$ spinors follow after substitution ${\bf p}
\rightarrow -{\bf p}$ in the matrix structures of the equations ({\it not}
in the spinors!). The phase factors
$\chi_{_{R,L}} \equiv \vartheta_1^{^{R,L}} +\vartheta_2^{^{R,L}}$
are defined by explicit forms of
the 2-spinors of different helicities\footnote{Of course, different
choices of $\chi_{_{R,L}}$ will have influence
Eqs. (\ref{i1},\ref{i2}).} and can be regarded at this moment as
arbitrary.

Considering properties of 4-spinors with respect to the ${\bf p}
\rightarrow -{\bf p}$ after S. Bruce we state\footnote{We prefer to use
the conventional  notation ${\bf M} \rightarrow {\bf E}$, the polar vector,
and ${\bf N} \rightarrow {\bf B}$, the axial vector. We  still leave a
room for different interpretations of these vectors in physical relevant
cases.}
\begin{equation}
\xi^{^L}_{even}=\phi_{_L}^\downarrow + \Theta \phi_{_L}^{\uparrow\,\,\ast}
\equiv \pmatrix{B^3 +i B^0\cr B^1 +i B^2}\quad,\quad
\xi^{^L}_{odd}=\phi_{_L}^\downarrow - \Theta \phi_{_L}^{\uparrow\,\,\ast}
\equiv \pmatrix{-E^0 +i E^3\cr -E^2 +i E^1}\quad.
\end{equation}
The  opposite helicity parts are connected by the Wigner time-reversal
operator:
\begin{equation}
\phi_{_L}^\uparrow + \Theta
\phi_{_L}^{\downarrow\,\,\ast} \equiv \Theta {\cal K} \xi^{^L}_{odd} =
\pmatrix{E^2 +i E^1\cr -E^0 -i E^3}\quad,\quad
\phi_{_L}^\uparrow - \Theta
\phi_{_L}^{\downarrow\,\,\ast} \equiv -\Theta {\cal K} \xi^{^L}_{even}=
\pmatrix{B^1 -i B^2\cr -B^3 +i B^0}\quad.
\end{equation}

Adding and subtracting equations (\ref{111}-\ref{114})
we obtain
\begin{mathletters}
\begin{eqnarray}
\pmatrix{p^0& 0\cr 0&p^0} \pmatrix{E^2 +iE^1\cr -E^0 -iE^3}
&+&\pmatrix{p^3 & p^1 -ip^2\cr p^1+ip^2 &-p^3\cr}
\pmatrix{B^1 -iB^2\cr -B^3 +iB^0\cr} -\nonumber\\
&-&m\cos\chi\pmatrix{E^2 +iE^1\cr -E^0 -iE^3\cr} +im \sin\chi \pmatrix{B^1
-iB^2\cr -B^3 +iB^0\cr} =0\, ,\\
\pmatrix{p^0& 0\cr 0&p^0} \pmatrix{B^1 -iB^2\cr -B^3 +iB^0}
&+&\pmatrix{p^3 & p^1 -ip^2\cr p^1+ip^2 &-p^3\cr}
\pmatrix{E^2 +iE^1\cr -E^0 -iE^3\cr} +\nonumber\\
&+&m\cos\chi\pmatrix{B^1 -iB^2\cr -B^3 +iB^0\cr} -im \sin\chi \pmatrix{E^2
+iE^1\cr -E^0 -iE^3\cr} =0\, ,\\
\pmatrix{p^0& 0\cr 0&p^0} \pmatrix{B^3 +iB^0\cr B^1 +iB^2}
&+&\pmatrix{p^3 & p^1 -ip^2\cr p^1+ip^2 &-p^3\cr}
\pmatrix{-E^0 +iE^3\cr -E^2 +iE^1\cr} +\nonumber\\
&+&m\cos\chi\pmatrix{B^3 +iB^0\cr B^1 +iB^2\cr} -im \sin\chi \pmatrix{-E^0
+iE^3\cr -E^2 +iE^1\cr} =0\, ,\\
\pmatrix{p^0& 0\cr 0&p^0} \pmatrix{-E^0 +iE^3\cr -E^2 +iE^1}
&+&\pmatrix{p^3 & p^1 -ip^2\cr p^1+ip^2 &-p^3\cr}
\pmatrix{B^3 +iB^0\cr B^1 +iB^2\cr} -\nonumber\\
&-&m\cos\chi\pmatrix{-E^0 +iE^3\cr -E^2 +iE^1\cr} +im \sin\chi \pmatrix{B^3
+iB^0\cr B^1 +iB^2\cr} =0\,\, .
\end{eqnarray} \end{mathletters}
They recast into the vector form:
\begin{mathletters}\begin{eqnarray}
{\bf p}\times {\bf E} -p^0 {\bf B} +{\bf p} E^0 -m  {\bf B} \cos \chi
-m {\bf E}\sin\chi &=& 0\quad,\\
{\bf p}\times {\bf B} +p^0 {\bf E} +{\bf p} B^0 -m {\bf E} \cos\chi
+m  {\bf B}\sin\chi &=& 0\quad,\\
p^0 E^0 - ({\bf p}\cdot {\bf B}) -m E^0 \cos\chi + m B^0 \sin\chi&=&
0\quad,\\
p^0 B^0 + ({\bf p}\cdot {\bf E}) +m B^0 \cos\chi+ m E^0 \sin\chi&=&
0\quad.
\end{eqnarray} \end{mathletters}
For parity conservation of these vector equations we
should assume that $E_0$ would be a pseudoscalar and $B_0$ would be a
scalar, furthermore, $\chi = 0\quad \mbox{or} \quad\pi$. In a matrix form
with the Majorana-Oppenheimer matrices
\begin{mathletters}
\begin{eqnarray}
\alpha^0 &=& \openone_{4\times 4}\quad,\qquad\qquad \alpha^1
=\pmatrix{0&-1&0&0\cr -1&0&0&0\cr 0&0&0&-i\cr 0&0&i&0\cr}\quad,\\ \alpha^2
&=&\pmatrix{0&0&-1&0\cr 0&0&0&i\cr -1&0&0&0\cr 0&-i&0&0\cr}\quad,\quad
\alpha^3 =\pmatrix{0&0&0&-1\cr 0&0&-i&0\cr 0&i&0&0\cr -1&0&0&0\cr}\quad
\end{eqnarray}
\end{mathletters}
we can obtain ($\overline{\alpha}^0 \equiv\alpha^0$\,\, ,\,\,
$\overline{\alpha}^i \equiv -\alpha^i$)
\begin{mathletters}
\begin{eqnarray}
\alpha^\mu p_\mu \Psi_2 (p^\mu) -me^{-i\chi}\Psi_1 (p^\mu)&=&0\quad,\\
\overline{\alpha}^\mu p_\mu \Psi_1 (p^\mu) -me^{+i\chi}\Psi_2
(p^\mu)&=&0\quad
\end{eqnarray} \end{mathletters}
for the field functions
\begin{mathletters}
\begin{eqnarray}
\Psi_1 (p^\mu) = -{\cal C} \Psi_2^\ast (p^\mu) = \pmatrix{-i (E^0-iB^0)\cr
E^1 -iB^1\cr
E^2-iB^2\cr
E^3 -iB^3\cr}\,\, ,\,\,
\Psi_2 (p^\mu) = -{\cal C} \Psi_1^\ast (p^\mu) = \pmatrix{-i (E^0+iB^0)\cr
E^1 +i B^1\cr
E^2+i B^2\cr
E^3 +i B^3\cr}\,\, ,
\end{eqnarray}
\end{mathletters}
where
\begin{equation}
{\cal C} = {\cal C}^{-1} =\pmatrix{1&0&0&0\cr
0&-1&0&0\cr
0&0&-1&0\cr
0&0&0&-1\cr}\quad,\quad {\cal C}\alpha^\mu {\cal C}^{-1} =
\overline{\alpha}^{\mu\,\,\ast}
\end{equation}
in accordance with the definition of Dowker~\cite{Dowker} in the
orthogonal basis.  Some remarks have already been done that these
equations can be written in the same form for both  positive- and
negative-frequency solutions.  If choose $\Psi_2 (p^\mu)$ as presenting a
field operator and $\chi=\pi$ then we have in the coordinate
representation:
\begin{mathletters} \begin{eqnarray}
i\wp_{u,v} \alpha^\mu \partial_\mu \Psi (x^\mu) - m \Psi^c
(x^\mu) &=&0 \label{mw1}\\
i\wp_{u,v} \overline{\alpha}^\mu \partial_\mu \Psi^c (x^\mu)
- m \Psi (x^\mu) &=&0 \label{mw2}
\end{eqnarray} \end{mathletters}
with
$\wp_{u,v} = \pm 1$ depending on what solutions, of positive or negative
frequency, are considered (cf.  with~\cite{BWW}).

Next, we would like to mention that similar formulations (but without
a mass term) were met in literature~\cite{Maj1,Oppen,Imaeda,Ohmura}.
Probably, applying  them was caused by some shortcomings of
the equation (1) of the paper~\cite{Bruce}, which the author of
the commented work paid attention to.  To his critical remarks we can
add that it contains the acausal solution with
$E=0$, ref.~\cite{Maj1,Oppen,Ahl1}. Acausal solutions
of the similar nature appear for any
spin (not only for spin-1 equations), ref.~\cite{Ahl1}.
While Oppenheimer proposed a physical interpretation of this solution as
connected with electrostatic solution and recently
another solution, the $B(3)$ longitudinal field,  to the Maxwell's
equations was extensively discussed~\cite{Evans} the problem did not yet
find an adequate consideration. In the mean time, the equations for spin-1
massless bosons, presented by Majorana, Oppenheimer, Giantetto, and the
ones of this paper for massive spin-1 case, are free of any acausalities;
they are of the first order in time derivatives and represent the
Lorentz-invariant theory.\footnote{The question of the relativistic
invariance of the equations (\ref{mw1},\ref{mw2}) is a tune point and due
to volume restrictions for {\it Note Brevi} we don't deal now with these
matters in detail. The separate paper will discuss the
relativistic invariance of new equations. But, one should note here
that providing new frameworks we are not going to dispute results of
the Dowker's consideration~[7a, p.183].} As a price
we have additional displacement current and a possible mass term.

Another equations which can be considered as suitable
candidates for describing spin-1 bosons are the second-order Weinberg
equations~\cite{Wein,Ahl1,Dvo942}; they have only causal solutions $E=\pm
p$ in the massless limit. Moreover, their massless limit also can be
reduced~\cite{Dvo942} to Eqs.  (24) of the commented paper.

Finally, I would like to note that  presented ideas deserve further
rigorous elaboration, since we are still far from understanding the nature
of electron, photon and neutrino. Their specific features seem not to lie
in some specific representation of the Poincar\`e group but  in the
structure of our space-time. Thus,  the equations for the fields in the
$(1,0)\oplus (0,0)$ and the $(0,1)\oplus (0,0)$ representations, given
above, could provide additional information for our goals.

\bigskip

{\it Acknowledgments.} I appreciate encouragements and discussions with
Profs. D. V. Ahluwalia, M. W. Evans, I. G. Kaplan and A.~F. Pashkov. Many
internet communications from colleagues are acknowledged.  I am grateful
to Zacatecas University for a professorship.  This work has been partially
supported by el Mexican Sistema Nacional de Investigadores, el
Programa de Apoyo a la Carrera Docente and by the CONACyT under the
research project 0270P-E.


\begin{references}

\footnotesize{
\baselineskip13pt

\bibitem{Bruce} S. Bruce, Nuovo Cimento B{\bf 110} (1995) 115

\bibitem{NP} D. V. Ahluwalia, Int. J. Mod. Phys. A{\bf 11} (1996) 1855;
V. V. Dvoeglazov, Rev. Mex. Fis. Suppl. {\bf 41} (1995) 159; Int. J.
Theor.  Phys.  {\bf 34} (1995) 2467; Nuovo Cim. A{\bf 108} (1995) 1467

\bibitem{Maj} E.  Majorana, Nuovo Cim. {\bf
14} (1937) 171 [English translation: Tech.  Trans. TT-542, Nat. Res.
Council of Canada]

\bibitem{MLC}  J. A. McLennan,  Phys. Rev. {\bf 106} (1957) 821;
K. M. Case, Phys. Rev. {\bf 107} (1957) 307

\bibitem{Ziino} A. O. Barut and G. Ziino, Mod. Phys. Lett.
A{\bf 8} (1993) 1011; G. Ziino, Int. J. Mod. Phys. A{\bf 11} (1996) 2081

\bibitem{Markov}  M. A.  Markov, ZhETF {\bf 7} (1937) 603 (see also
ref. there: Gehenian, Compt. Rend. {\bf 198} (1934) No. 8)

\bibitem{Dowker} J. S. Dowker and Y. P. Dowker, Proc. Roy. Soc. A{\bf
294} (1966) 175; J.  S.  Dowker, ibid  A{\bf 297} (1967) 351

\bibitem{BWW} D. V. Ahluwalia, M. B. Johnson and T. Goldman,
Phys. Lett. B{\bf 316} (1993) 102. This paper presents an explicit example
of the quantum field theory of the Bargmann-Wightman-Wigner-type
[E. P. Wigner, en {\it Group Theoretical Concepts and
Methods in Elementary Particle Physics (Lectures of the Istanbul Summer
School of Theoretical Physics, 1962, Ed. F. G\"ursey)}, Gordon \& Breach,
1965, p. 37]

\bibitem{Maj1} E. Majorana, Scientific Manuscripts (1928-1932),
edited by R. Mignani, E. Recami and M. Baldo, Lett. Nuovo Cim. {\bf 11}
(1974) 568; see also E. Gianetto, Lett. Nuovo Cim. {\bf 44} (1985) 140

\bibitem{Oppen} J. R. Oppenheimer, Phys. Rev. {\bf  38}
(1931) 725

\bibitem{Imaeda} K. Imaeda, Prog. Theor. Phys. {\bf 5} (1950) 133

\bibitem{Ohmura} T. Ohmura (Kikuta), Prog. Theor. Phys. {\bf 16} (1956)
684, 685

\bibitem{Ahl1} D. V. Ahluwalia and D. J. Ernst, Mod. Phys. Lett.
A{\bf 7} (1992) 1967

\bibitem{Evans} M. W. Evans and J.-P. Vigier, {\it Enigmatic Photon.} Vol.
1 \& 2 (Kluwer Academic Pub., Dordrecht, 1994-95);
see also V. V. Dvoeglazov Yu. N. Tyukhtyaev and S. V.  Khudyakov, Russ.
Phys. J. {\bf 37} (1994) 898

\bibitem{Wein} S. Weinberg, Phys. Rev. B{\bf 133} (1964) 1318

\bibitem{Dvo942} V. V. Dvoeglazov, {\it Can the $2(2j+1)$
Component Weinberg-Tucker-Hammer Equations Describe the Electromagnetic
Field?} Preprint hep-th/9410174, Zacatecas, Oct. 1994

}

\end{references}
\end{document}